\documentclass{svproc}
\usepackage{graphicx}%
\usepackage{multirow}%
\usepackage{amsmath,amssymb,amsfonts}%
\usepackage{mathrsfs}%
\usepackage[title]{appendix}%
\usepackage{xcolor}%
\usepackage{textcomp}%
\usepackage{manyfoot}%
\usepackage{booktabs}%
\usepackage{algorithm}%
\usepackage{algorithmicx}%
\usepackage{algpseudocode}%
\usepackage{listings}%
\usepackage{url}
\usepackage{glossaries}
\usepackage{glossaries-extra}
\usepackage{subcaption}
\usepackage[frozencache]{minted}
\usepackage{enumitem}
\usepackage{booktabs}
\usepackage{microtype}

\setabbreviationstyle[acronym]{short}
\newacronym{cg}{CG}{Call Graph}
\newacronym{bb}{BB}{Basic Block}
\newacronym{cfg}{CFG}{Control Flow Graph}
\newacronym{gnn}{GNN}{Graph Neural Networks}
\newacronym{ml}{ML}{Machine Learning}
\newacronym{dl}{DL}{Deep Learning}
\newacronym{nlp}{NLP}{Natural Language Processing}
\newacronym{cff}{CFF}{CFG Flattening}
\newacronym{mba}{MBA}{Mixed Boolean Arithmetic}
\newacronym{ha}{HA}{Hungarian Algorithm}
\newacronym{gmn}{GMN}{Graph Matching Networks}
\newacronym{sog}{SOG}{Semantic Oriented Graph}
\newacronym{gcn}{GCN}{Graph Convolutional Network}
\newacronym{lstm}{LSTM}{Long Short-Term Memory}
\newacronym{ir}{IR}{Intermediary Representation}
\begin{document}
\mainmatter              % start of a contribution
\title{Identifying Obfuscated Code through Graph-Based Semantic Analysis of Binary Code}
\titlerunning{Identifying Obfuscated Code through Graph-Based Semantic Analysis}  % abbreviated title (for running head)
\author{Roxane Cohen\inst{1, 2} \and Robin David\inst{1} \and 
Florian Yger\inst{3} \and Fabrice Rossi\inst{4}}
\authorrunning{Cohen et al.} % abbreviated author list (for running head)

\institute{Quarkslab,\\
%\email{rcohen@quarkslab.com},\\
\and
LAMSADE, CNRS, Université Paris-Dauphine - PSL, Paris, France\\
\and 
LITIS, INSA Rouen Normandy, Rouen, France\\
\and
CEREMADE, CNRS, Université Paris-Dauphine - PSL, Paris, France}

\maketitle              % typeset the title of the contribution

\begin{abstract}
Protecting sensitive program content is a critical issue in various
situations, ranging from legitimate use cases to unethical contexts. 
Obfuscation is one of the most used techniques to ensure such
protection. Consequently, attackers must first detect and characterize
obfuscation before launching any attack against it. 
This paper investigates the problem of function-level obfuscation
detection using graph-based approaches, comparing algorithms, from
elementary baselines to promising techniques like \glsfmtfull{gnn}, on
different feature choices. We consider various
obfuscation types and obfuscators, resulting in two complex datasets. Our findings demonstrate that \glspl*{gnn} need
meaningful features that capture aspects of function semantics to
outperform baselines. Our approach shows satisfactory results, 
especially in a challenging 11-class classification task and in a practical malware analysis example.
%two practical binary analysis examples.  
\end{abstract}
\section{Introduction} \label{sec1}
Binary programs and their sensitive contents are often protected from reverse engineering through
obfuscation techniques \cite{Collberg2009Book}, which aim to obscure a
program's underlying logic without altering its functionality. While
reverse engineers seek to comprehensively understand program
semantics \cite{Raja2007RE}, developers try to conceal them, at least partially. Their
motivations range from legitimate concerns like intellectual property
protection to less ethical practices such as hiding malicious
payloads \cite{Sharif2008Malware}. Consequently, detecting obfuscated programs is useful for
program protection assessment and malware detection complementing, for instance, traditional
signature-based methods. \glsfmtfull{ml} approaches have emerged as
effective tools for this detection task \cite{Greco2023Explaining}.

In this paper, we focus on obfuscation detection at the function
level, with the aim of identifying both obfuscated functions and the
specific obfuscation techniques employed. Our objective is to pinpoint
obfuscated functions within a binary. While this can help determine if
a program is obfuscated overall \cite{Greco2023Explaining}, it has broader implications. First, as obfuscation negatively impacts program
efficiency, developers tend to only obfuscate important functions, which ultimately are the ones of interest for an analyst. 
Second, automated attacks have been developed against
specific obfuscation schemes, assuming the obfuscation is already detected and located \cite{David2020DeobfuscationMBA,Yadegari2015DeobfuscationGeneral,David2017XTunnel,Salwan2018DeobfuVirtualize,Ramtine2019Opaque}. By detecting functions obfuscated with
these schemes, analysts can effectively employ the corresponding
attacks. Finally, it provides an evaluation of how stealthy an obfuscation is, which is crucial for
developers, as automated detection tools can provide a measure of the
undetectability of obfuscation methods at a fine-grained level \cite{Kanzaki2015Artificiality}. %,Blazytko2023StatAnalysis}. 

This problem has been studied from a \gls*{ml} perspective in
several previous works. Many of these studies extract various features
from binary programs, such as the distribution of instructions in the
assembly code \cite{Salem2016TFIDF}, the function complexity metrics
\cite{Schrittwieser2023Stealth} or semantic reasoning \cite{Ramtine2019Stealth}. Some of the
more advanced features are derived from a graph-based representation of
the binary functions. Specifically, each function can be represented by its
\glsfmtfull{cfg}, a graph where each node is a \glsfmtfull{bb}, an atomic
sequence of instructions without any branching and where a
directed edge between two nodes corresponds to a candidate execution path. The cyclomatic complexity is one example of
a feature computed from the \gls*{cfg}.

While extracting domain-specific features from the \gls*{cfg} has been
proved to be effective in some cases \cite{Greco2023Explaining},
leveraging the full structural information of the graph can enhance
classification performance in certain applications. This can be
achieved using kernel methods \cite{Kriege2020} or
\glsfmtfull{gnn}. \glspl*{gnn} are under very active development since their
introduction \cite{Gori2005GNN} and have shown promise in
outperforming feature-based approaches in specific scenarios
\cite{Errica2020A}.

Function-level obfuscation
detection and classification were investigated using a specific type of \gls*{gnn}, a
\glsfmtfull{gcn} combined with a 
\glsfmtfull{lstm} neural network
\cite{Jiang2021GCNLSTM}. This approach outperformed baseline methods
based on manually extracted features. However, these features were limited to the \gls*{bb} level and
combined using a simple sum, lacking structural features that could be derived from
the \gls*{cfg}. 

In this paper, we compare function-level C code obfuscation detection and
classification methods based on advanced features, including
structural features extracted from the \glspl*{cfg}, processed by
classical \gls*{ml} algorithms (Random Forest and
Gradient Boosting), to methods based on \glspl*{gnn} that
directly process attributed \glspl*{cfg}. We extend the previous \gls*{gnn} approach
\cite{Jiang2021GCNLSTM} by comparing different collections of
features, including graph-level ones, and exploring various
\glspl*{gnn} architectures. We investigate more advanced obfuscation
techniques than those provided by OLLVM \cite{ollvm} by incorporating
Tigress \cite{tigress} in our experiments. We use a larger dataset and investigate two data splits to control the
classification difficulty. Our experiments demonstrate that obfuscation detection is best achieved through a \gls*{gnn} processing of fine-grained semantic level features.

%JavaScript and Android obfuscations need specific detection techniques \cite{Sheneamer2018Java}. 

The remainder of this paper is organized as follows. Section
\ref{sec:obfuscation} introduces the obfuscation techniques considered
in this study and briefly discusses their impact on \gls*{cfg}. Section \ref{graphreplearning} reviews fundamental concepts
related to \glspl*{gnn}. Section \ref{sec:dataset} describes
the proposed dataset and Section \ref{sec:methods} 
outlines our experimental setup. Section
\ref{sec:bin} presents the initial task of binary classification,
followed by the extended multi-class experiment in Section
\ref{sec:mul}. Section \ref{sec:xtunnel} shows a real-world malware example dedicated to obfuscation detection. A discussion in Section \ref{conclusion} concludes this study. 

\section{Binary representation and obfuscation} \label{sec:obfuscation}

Binary code is often described by its corresponding disassembly, a symbolic representation of the machine code. At function level, this disassembly is naturally represented with an attributed \gls{cfg} that details the function execution flow between atomic blocks of code, denoted as \gls{bb}. Such a representation is particularly useful as semantic information can be extracted from it, describing the function behavior.

However, binary code is extremely sensitive to compilation parameters, such as the optimization level. A given function can have multiple \glspl*{cfg} representations that all convey the same underlying semantics. Conversely, two different functions may share the same \gls*{cfg} structure but differ on the instructions in their \glspl*{bb}. 

A different source of variability is induced by program
obfuscation. It aims at altering a program syntax but
not its behavior. It consists in specific transformation passes that try to increase code security against reverse engineering. Obfuscation is widely used to protect binary assets, such as data, keys or algorithms. 
%Obfuscation samples can be directly linked with adversarial examples, as they are conceived to hide their true nature, by making them looking completely different from what they really are. 
Each obfuscation pass has specific effects on binaries
\cite{Collberg2009Book}. In particular, a \textbf{data-related obfuscation} consists in modifying the
function data-flow. For example, a \glsfmtfull{mba} \cite{Zhou2007MBADef} replaces
integer values with a sequence of complex arithmetic computations that
is strictly equivalent. A \textbf{control-flow obfuscation} modifies
the true program execution flow, either at the function level or at
the program level. One elementary obfuscation among this type is the
\glsfmtfull{cff}, that, inside the function, puts every \gls*{bb} at
the same level and uses a dispatcher to preserve the execution flow
logic \cite{Wang2001CFFDef}. 
 
Figure \ref{img_src} and Table \ref{img_opti_levels} illustrate the high variability of binary code, depending on the compilation effect or obfuscation. Intuitively, detecting a pass that has a subtle effect on binary code, such as \gls*{mba}, is tedious. The resulting function may be confused with a legitimate complex one with dozens of arithmetic operations.

\begin{figure}
    \begin{minted}[frame=lines, fontsize=\scriptsize, linenos, bgcolor=white]{c}
const char * ZEXPORT gzerror(gzFile file, int *errnum) {
    gz_statep state;

    if (file == NULL)
        return NULL;
    state = (gz_statep)file;
    if (state->mode != GZ_READ && state->mode != GZ_WRITE)
        return NULL;

    if (errnum != NULL)
        *errnum = state->err;
    return state->err == Z_MEM_ERROR ? "out of memory" :
                    (state->msg == NULL ? "" : state->msg);
}
    \end{minted}
    \vspace{-10mm}
    \caption{\texttt{gzerror} function source code (\texttt{zlib} project)}\label{img_src}
\end{figure}
\begin{table}[ht!]
    % \begin{center}
    \begin{tabular}{ c | c | c }
        \toprule
        -O0 optimization & -O2 optimization & Obfuscated with \gls*{cff} \\ 
        \cmidrule(r){1-1}\cmidrule(lr){2-2}\cmidrule(l){3-3}
        \raisebox{-\totalheight}{\includegraphics[width=0.25\textwidth, height=35mm]{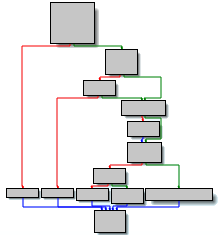}}
        & 
        \raisebox{-\totalheight}{\includegraphics[width=0.25\textwidth, height=35mm]{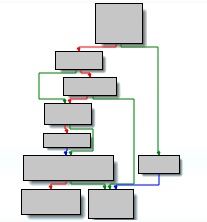}}
        & 
        \raisebox{-1.1\totalheight}{\includegraphics[width=0.46\textwidth, height=30mm]{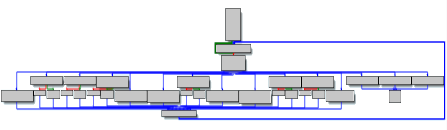}}
        \\ \bottomrule
     \end{tabular}
     \caption{Optimization and obfuscation effects on the \texttt{gzerror} function \gls*{cfg}.}
     \label{img_opti_levels}
    %  \end{center}
\end{table}

\section{Machine learning for graphs} \label{graphreplearning}

Graph representation learning aims at encoding graph data into a low-dimensional vector. There exist two main classes of algorithms for this task: feature-based approaches and \glspl*{gnn}. 
%\item Walk-based algorithms
Feature-based approaches consist in using various expertly designed features to describe a graph, that are then often processed by traditional \gls*{ml} algorithms, such as Random Forest, for classification purpose \cite{Errica2020A}. Standard features are the number of nodes and edges, the mean node degree, the density, etc.
%Walk-based methods propose to iteratively visit a graph by randomly exploring its nodes and edges. Then, random walks can describe the graph structure and Node2vec \cite{Grover2016Node2vec} is the best example of such an approach.

\glsfmtfull{gnn} have experienced recent popularity, despite having been theorized quite early \cite{Gori2005GNN}. \glspl*{gnn} use graph structure and initial features assigned to the graph nodes to iteratively learn either node or graph representation, with a message passing mechanism. Formally, the $k$-th layer of a message passing \gls*{gnn} is described as follows \cite{Xu2019Howpowerful}:
\begin{align*}
a_v ^{(k)} &= AGG^{(k)} \left( \{ h_u^{(k-1)} : u \in \mathcal{N}(v) \} \right),&
h_v^{(k)} &= COMB ^{(k)} \left(h_v^{(k-1)}, a_v^{(k)} \right), 
\end{align*}
where $h_v^{(k)}$ is the feature associated to the node $v$ at the $k$-th iteration, $h_v^{(0)} = X_v$ is the initial feature of node $v$ and $\mathcal{N}(v)$ denotes the neighborhood of node $v$. Such node representation can be directly used for node-level tasks. For graph-level tasks, a graph embedding can be derived using a readout function that will combine all the node representations into a final graph vector:
\begin{equation*}
h_G = RO ( \{ h_v^{(K)} | v \in G \} )
\end{equation*}

The AGG, COMB and the optional RO functions vary depending on the
message passing \gls*{gnn} model that is used. GCN
\cite{Welling2017GCN} was the first applicable \gls*{gnn} using
convolutional layers. SAGE \cite{Hamilton2018SAGE} is a refined GCN
version, with a more advanced COMB method. GIN is a model architecture
that offers the best theoretical foundations as it has been shown to
be as powerful as the 1-Weisfeiler-Lehman test
\cite{Xu2019Howpowerful}. GAT includes an attention mechanism in the
message passing framework, that should give more weight to important
nodes \cite{Velickovic2018GAT}. UNet is inspired from the usual UNet
architecture for computer vision, where the dimension of the input
data is first downsampled and then expended again \cite{Gao2019Unet}.

As mentioned before, \glspl*{gnn} take as input a graph with $n$ nodes and a feature matrix of dimension $(n, d)$ with $d$, the feature dimension. This node feature matrix should ideally describe semantically the content or the nature of graph nodes. Contrarily to feature-based approaches, \glspl*{gnn} use node-level features. They are known to have a huge impact during the \gls*{gnn} training \cite{Errica2020A}. 

Beyond the \gls*{gnn} popularity, one must remember that simple baselines should always be used as a comparison. Previous works highlight the fact \glspl*{gnn} do not always provide the best results compared to baselines that are less costly \cite{Errica2020A,Ferrari2019Progress}.

\section{Dataset}\label{sec:dataset}
In order to study obfuscation, we have built an open-source dataset
\cite{obfdataset} that consists of C program sources obfuscated by two
different obfuscators and with different obfuscation types
(control-flow and data obfuscations). The
dataset is based on five open source projects: \texttt{zlib},
\texttt{lz4}, \texttt{minilua}, \texttt{sqlite} and \texttt{freetype},
compiled for x86-64, that are obfuscated with
Tigress \cite{tigress}, a source-to-source obfuscator, and OLLVM
\cite{ollvm}, that directly interfaces with the compiler. Several
Tigress obfuscations are selected: 
\begin{itemize}
\item Data obfuscation: EncodeArithmetic, EncodeLiterals;
\item Control-flow obfuscation: Virtualize, OpaquePredicates, \gls*{cff}, Split, Merge, Copy;
\item Combined: a mix of \gls*{cff}, EncodeArithmetic and OpaquePredicates, and the same mix with an additional Split.
\end{itemize}
OLLVM unfortunately offers less obfuscations: only OpaquePredicates,
\gls*{cff} and a pass similar to EncodeArithmetic.  

Data leakage can be a serious issue when building such a dataset. A
particular risk is that an obfuscation could be negated by compiler optimizations. 
Another potential issue is functions shared by
different projects. To avoid any leakage, we use two data split
strategies to produce a training set, a validation set and a test set.

In the \textbf{per function split strategy} (\textit{Dataset-1}), a
function and its obfuscated versions are within the same set. To implement
this strategy, all the functions of the five projects are collected,
ensuring the common functions between projects, such as the
\texttt{libc} functions, are completely removed. The resulting
function list is randomly split into three sets: train, validation and
test with a ratio of (64\%, 16\%, 20\%) of the functions, using
stratified sampling to ensure a similar distribution of \gls*{bb} number
across sets. For each function in a subset, we include both the obfuscated and unobfuscated versions. Notice that this
leads to an unbalanced class ratio, as the original dataset contains
11 obfuscation classes for only one unobfuscated class. This class
unbalance ratio is unusual in a context of anomaly detection, as in general there
exist much more normal data than abnormal ones. However, obtaining
abnormal data is accessible in this context, even if in practice,
obfuscated functions (abnormal data) are often limited inside a binary
for computational reasons.

In the \textbf{per binary split strategy} (\textit{Dataset-2}), we use
all the functions belonging to \texttt{zlib, lz4} and
\texttt{minilua}, and all their obfuscated versions to create
the train and validation sets. These two sets are split
according to the same procedure used to generate the \textit{Dataset-1}, with a
ratio of (80\%, 20\%). The test set is made of all the functions of
\texttt{sqlite} and \texttt{freetype}, with all their obfuscated
versions. This setting represents a real-world scenario for detecting
obfuscation: we want to detect and characterize obfuscated functions 
in a completely new executable using a model trained on controlled binaries that are potentially 
unrelated to the new one.

\textit{Dataset-2} should be more challenging than \textit{Dataset-1} as two projects
may have a different coding style or may use a vastly different number
of functions that are valid candidates for certain types of
obfuscation. Having different projects during training/validation and
testing prevents the model from leveraging per project regularity. 

For these two datasets, -O0 and -O2 binaries are separated. Indeed,
compiler optimizations tend to remove, sometimes completely, the
applied obfuscation. This is particularly true for the OLLVM
obfuscator, leading to many obfuscated variants that are identical to
non-obfuscated versions. The dataset descriptions, especially the number of
functions and samples, and the class ratio are available in Table
\ref{datasets}.  

\begin{table}[t]
    \centering
    \caption{Characteristics of the two datasets}
    \begin{tabular}{|c|rl|rl|}
        \hline
        & \multicolumn{2}{c|}{\bf-O0 optimization} & \multicolumn{2}{c|}{\bf-O2 optimization} \\
        \hline
        \multirow{5}{*}{\bf \textit{Dataset-1}} &
            Train* :      & 3,225 / 48,813    & Train* :      & 1,846 / 23,151 \\
          & Validation* : & 803 / 12,135      & Validation* : & 459 / 5,753 \\
          & Test* :       & 1012 / 15,403     & Test* :       & 583 / 7,162 \\
    & Ratio binary=                        & (0.11, 0.11, 0.11)     & Ratio binary:                        & (0.17, 0.17, 0.17) \\
          %& Ratio multi-class=                   & ()                     & Ratio multi-class:                   & () \\
          %& p-value=                             & 0.99                   & p-value=                             & 0.98 \\

        \hline
        \multirow{5}{*}{\bf \textit{Dataset-2}} &
                Train* :       & 1,137 / 18,759   & Train* :      & 610 / 9,019  \\
              & Validation* :  & 279 / 4,652       & Validation* : & 150 / 2,238 \\
              & Test* :        & 3,948 / 57,627    & Test* :       & 3012 / 31,760 \\
        & Ratio binary : & (0.13, 0.13, 0.11)& Ratio binary : & (0.14, 0.14, 0.17)\\
              %& Ratio multi-class : & ()           & Ratio multi-class : & () \\
              %& p-value =      & 0.98              & p-value =     & 0.97 \\
        \hline
        \multicolumn{5}{l}{* values expressed in functions/samples}
    \end{tabular}
    \label{datasets}
\end{table}

\section{Methods and feature vectors}\label{sec:methods}

Quantifying precisely to what extent each aspect of the obfuscation detection problem contributes to the final classification score is essential. 
%Indeed, if the \gls*{cfg} represents the first intuitive graph for a function, more graph construction can be obtained. 
In fact, various features are potential candidates to be used for further classification: textual data from assembly text, statistical data, etc. In this work, graph representation and features are gradually enriched, starting from existing \gls*{ml} works, before diving into more advanced \gls*{gnn} algorithms. 

All our experiments were conducted on a Linux-based server equipped of Nvidia RTX A6000, with 20 cores, 40 threads and 32GiB of RAM.

%\subsubsection{Graph types}

%Several types of graphs are considered: 
%\begin{itemize}
%\item \gls*{cfg}, where nodes represent \glspl*{bb} and edges denote the execution flow between these nodes inside the function. This graph is attributed and each node contains its associated assembly code. 
%\item \gls*{cfg}-IR. This graph is based on the elementary \gls*{cfg}, except it uses an intermediary representation instead of the raw assembly code. Such intermediary representation is based on Pcode \cite{pypcode}. Each assembly instruction is equivalent to a list of more abstract Pcode instructions. Then, it is possible to create a \gls*{cfg}-IR, where each node corresponds to an atomic Pcode block of code, linked by Pcode execution flow. Such \gls*{cfg}-IR is not expected to differ too significantly from the initial \gls*{cfg}: nodes contains more instructions but the general topology of the graph is preserved\footnote{Specific assembly instructions inside a \gls*{bb} can be translated as a list of Pcode instructions that contains a branching one (which means the block is no longer atomic). When it happens, the structure of the \gls*{cfg} and \gls*{cfg}-IR differs, leading to more edge in the \gls*{cfg}-IR graph.}. Using \gls*{cfg}-IR representation gives rid of the noise specific to assembly code. It helps to be architecture-agnostic and more resilient to compiler optimization effects. 
%\end{itemize}

This study analyzes various graph-related algorithms. Indeed,
\gls*{cfg} are an intuitive graph representation for a binary
function, where nodes represent \glspl*{bb} and edges denote the
execution flow between these nodes inside the function. This graph is
attributed and each node contains its associated assembly code,
a sequence of instructions. Each instruction, e.g. \textit{mov eax, 0} in x86-64, combines a mnemonic
(the action to operate, here \textit{mov}) and operands (the arguments of this
action, here \textit{eax, 0})
%\footnote{The x86-64 instruction \textit{mov eax, 0} contains
%  the mnemonic \textit{mov} with operands \textit{eax, 0}. It assigns
%  the value 0 to the register \textit{eax}.}. 
As mentioned before, two
main algorithm classes are used for graph classification.

We use as baseline models traditional \gls*{ml} models, namely
\textbf{random forests} and \textbf{gradient boosting}. They operate
on \textbf{graph-level} features extracted from the functions. More precisely, we consider
two types of features. We extract first \gls*{cfg}-related features such as the number of nodes, edges, the cyclomatic
complexity, etc. They are complemented by \textbf{assembly TF-IDF
  features}. We use the approach proposed by Salem and
al. \cite{Salem2016TFIDF}, that is the counts of the
128 most used assembly mnemonics inversely weighted by the global
frequencies. 
%\footnote{The complete list is: the number of nodes, the number of edges, the number of strongly connected components, the cyclomatic complexity, the maximum number of instruction per node, the averaged number of instructions per node, the total number of instructions per function, the mean degree, density, diameter, transitivity of the \gls*{cfg}, its number of components, the number of \textit{IMUL}, \textit{SHR}, \textit{SHL}, \textit{SAR}, \textit{DIV}, \textit{XOR}, \textit{ADD}, \textit{SUB} instructions in the function, the number of Immediate in the function, the maximum degree and the minimum degree of the corresponding \gls*{cfg}. }

We compare these baselines to five \textbf{message passing \gls*{gnn}
  models}: GCN, SAGE, GIN, GAT and UNet. As mentioned in Section
\ref{graphreplearning}, \glspl*{gnn} need to have access to both graph
structure and node features that are given by the user. These \textbf{node-level} features are distinct to the ones of previous baselines that directly handle graph-level features. These
node feature vectors are iteratively refined as follows.

As a reference, we use an \textbf{identity feature vector}, a one-dimensional
vector filled with 1's. This forces the \gls*{gnn} to rely only on the
graph structure.

The first non-trivial feature vector is based on \textbf{counting assembly
  mnemonic classes}. We adopt here an existing strategy
\cite{Jiang2021GCNLSTM} which provides a coarse representation of the
assembly mnemonic distribution per \gls*{bb}.

To provide a more robust and refined representation, we use
\textbf{semantic and counting Pcode mnemonic}. Pcode is an
\glsfmtfull{ir} for which each assembly instruction
is semantically represented by one or more Pcode instructions in an
architecture agnostic way. Consequently, all the CPU architectures (ARM, Aarch64, MIPS) share the same underlying Pcode. This feature vector combines several
\gls*{cfg} \gls*{bb} features such as the number of instructions per
node with the Pcode mnemonic counts.

Finally, we consider \textbf{semantic and counting assembly mnemonic} which is
similar to the previous feature vector but uses counts of all possible
assembly instructions (1,828 mnemonic counts for binaries compiled in
x86-64). Such a feature is specific to an architecture, contrary to Pcode. 

%\footnote{The complete list is: a boolean value indicating if the basic block is the first one of the function, label encoding depending of the last mnemonic of a node (conditional jump, unconditional jump, call, ret, other), the number of instruction of a node, the number of its successors and the number of its predecessors}, followed by a counting vector of all the Pcode mnemonics (73 at most \cite{pypcode}). This second feature contains more data, thus is more robust, by explicitly describing all Pcode operations. Besides, this Pcode-based feature can be used whatever the chosen architecture. 
%\Using \glsfmtfull{nlp}-inspired approach for sequence analysis on assembly code: obtaining a text-embedding feature, based on the assembly, that is then given to a \gls*{gnn} is considered as state-of-the-art for the binary similarity problem \cite{Massarelli2019Investigating}. Consequently, such an approach is evaluated by taking advantage of existing transformer models trained on x86-64 assembly code, like PalmTree \cite{Li2021Palmtree}.

The above models are evaluated with their corresponding candidate
feature vectors, on both \textit{Dataset-1} and \textit{Dataset-2}. To select the best hyperparameters on the validation set, GridSearchCV and Optuna \cite{optuna} are respectively used for the baselines and the \glspl*{gnn}. The Optuna search is applied with three seeds, with the best run leading to the chosen hyperparameters, such as the number of layers or the hidden dimensions. Each Optuna run is restricted to 20 trials in order to limit the computational burden. Baselines and \glspl*{gnn} are respectively implemented using scikit-learn \cite{scikit_learn} and Pytorch-Geometric \cite{torchgeometric}.

Because of the unbalanced classes, both in binary and multi-class settings, our benchmarks are evaluated using the balanced accuracy. This metric heavily penalizes cases where a class is not properly detected compared to the others.

In this work, results for -O2 are omitted for brevity since they are based on the same principles as -O0 and show the same trends.

\section{Binary classification} \label{sec:bin}

\begin{table}[tb]
\caption{Binary classification scores, depending on features, algorithms and datasets.}
\centering
\begin{tabular}{|c|l|c|c|c|}
\hline
    \multirow[c]{2}{*}{\bf Graph} &
    \multicolumn{1}{c|}{\multirow[c]{2}{*}{\bf Features}} &
    \multirow[c]{2}{*}{\bf Algorithm} &
    \multicolumn{2}{c|}{\bf Balanced accuracy}   \\
\cline{4-5}  
 & & & \bf \textit{Dataset-1} & \bf \textit{Dataset-2}\\ 
    \hline
\multirow[c]{28}{*}{CFG}           
            & Graph features \&           &  RandomForest & 0.702 & 0.60  \\  \cline{3-5} 
            & assembly (Dim: {\bf \#23})  &  GradientBoosting & 0.725 & 0.649    \\  \cline{2-5}
            & TF-IDF on assembly          & RandomForest & 0.76 & 0.607    \\  \cline{3-5} 
            & mnemonics (Dim: {\bf \#128})& GradientBoosting & 0.80 & 0.683  \\  \cline{2-5} 
            & \multirow[c]{5}{*}{Identity (Dim: {\bf \#1})}
            		    &  GCN & 0.634 & 0.608  \\  \cline{3-5}
            &           &  Sage & 0.615 & 0.574      \\  \cline{3-5}
            &           &  GIN & 0.603 & 0.531            \\  \cline{3-5}
            &           &  GAT & 0.589 & 0.539       \\  \cline{3-5}
            &           &  UNet & 0.616 & 0.555       \\ \cline{2-5}
            & \multirow[b]{3}{*}{Counting mnemonic}
            		    &  GCN & 0.659 & 0.658  \\  \cline{3-5}
            &           &  Sage & 0.694 & 0.66        \\  \cline{3-5}
            &                           &  GIN & 0.701 & 0.673        \\  \cline{3-5}
            & classes (Dim: {\bf \#27}) &  GAT & 0.655 & 0.667       \\  \cline{3-5}
            &           &  UNet & 0.66 & 0.654    \\ \cline{2-5}
            &
                                  &  GCN & 0.789 & 0.736      \\  \cline{3-5}
            & Semantic \& counting&  Sage & 0.801 & 0.755     \\  \cline{3-5}
            & PCode mnemonics     &  GIN & 0.80 & 0.766         \\  \cline{3-5}
            & (Dim: {\bf \#78})   &  GAT & 0.805 & 0.731        \\  \cline{3-5}
            &                     &  UNet & 0.779 & 0.672        \\ \cline{2-5}
            &
                                  &  GCN & 0.792 & 0.758          \\  \cline{3-5}
            & Semantic \& counting&  Sage & 0.802 & 0.727          \\  \cline{3-5}
            & assembly mnemonics       &  GIN & 0.793  & 0.727            \\  \cline{3-5}
            & (Dim: {\bf \#1839}) &  GAT & 0.797 & 0.729         \\  \cline{3-5}
            &                     &  UNet & 0.785 & 0.701       \\ \cline{1-5}
%            & \multirow[c]{3}{*}{\#128: Transform model (PalmTree) on assembly code}
%                        &  GCN &           \\  \cline{3-5}
%            &           &  Sage &           \\  \cline{3-5}
%            &           &  GIN &       \\  \cline{3-5}
%            &           &  GAT &          \\  \cline{3-5}
%            &           &  UNet  & \\ \cline{1-5}
%\multirow[c]{9}{*}{CFG-IR}           
%            & \multirow[c]{2}{*}{Graph-level features \& Pcode counting}
%                        &  RandomForest & 0.662 &      \\  \cline{3-5}
%            &           &  GradientBoosting & 0.68 &         \\  \cline{2-5}
%            & \multirow[c]{2}{*}{Pcode TF-IDF}
%                        & RandomForest &  0.709 &       \\  \cline{3-5}
%            &           & GradientBoosting & 0.739 &          \\  \cline{2-5}
%            & \multirow[c]{3}{*}{\#77: Semantic + counting PCode mnemonics}
%                        &  GCN & 0.784 &      \\  \cline{3-5}
%            &           &  Sage & 0.734 &        \\  \cline{3-5}
%            &           &  GIN & 0.769 &        \\  \cline{3-5}
%            &           &  GAT & 0.768 &        \\  \cline{3-5}
%            &           &  UNet & 0.776 &  \\ \cline{1-5}
\end{tabular}
\label{binary_classif_O0}
\end{table}

In this Section, we address a simplified binary classification
problem where the goal is simply to determine if a function is
obfuscated or not. Results for -O0 are available in Table
\ref{binary_classif_O0}.

We note first that baselines with graph-level features demonstrate
satisfactory results, with a balanced accuracy that is better for the
\textit{Dataset-1}. Such behavior is expected as the \textit{Dataset-2} framework is
more challenging. Gradient Boosting outperforms slightly
Random Forest. Besides, the TF-IDF baseline respectively reaches 0.80 and
0.68 of balanced 
accuracy for \textit{Dataset-1} and \textit{Dataset-2}. This highlights the fact that a
fine-grained representation of the assembly mnemonic distribution
characterizes better the abnormality induced by obfuscation than graph-level features with a coarse-grained assembly representation. 

We observe that enriching \gls*{gnn} initial features is fundamental:
elementary features, such as the identity, offer poor
performances compared to more accurate features. This is consistent with the inferior performances of structural graph features with baseline methods. Moreover, the
best performances are obtained with the richest 
representation.

%that relies on Pcode instructions. It presents the advantage of being applicable to any CPU architecture, not only x86-64.
%To compare the Pcode efficiency to that of assembly instructions, we repeat this experiment by counting instead the x86-64 assembly mnemonics, leading to a feature of size 1839. It does not improve classification scores, advocating for an extensive of Pcode representation.
Using a feature based on assembly mnemonic counts provides approximately the
same results than relying on the Pcode instructions, even though it is much larger and restricted to a specific
architecture, which is not the case for Pcode, applicable to any CPU architecture, not
only x86-64. This demonstrates the Pcode potential to extract semantic features. Besides, theoretical results \cite{Xu2019Howpowerful} about GIN expressivity
power are confirmed by these experiments as GIN is generally slightly
more efficient than other \gls{gnn}.

%\item Contrary to many state-of-the-art papers on binary similarity \cite{investigating,fusion,bindiffnn}, using first a model language trained on assembly binaires and use them to initialize \gls*{gnn} initial features shows disappointing performances and is not sufficient to boost \gls*{gnn} score up to the baselines.
%\item Results for the CFG-IR show similar f1-score values than \gls*{cfg} representation. If baselines are less efficient in this setup, performances for \gls*{gnn} are not affected. Such graph representation can be generalized to any type of binary architecture. 

\begin{table}[tb]
\caption{Multi-class classification scores, depending on features, algorithms and datasets.}
\centering
\begin{tabular}{|c|l|c|c|c|}
\hline
    \multirow[c]{2}{*}{\bf Graph} &
    \multicolumn{1}{c|}{\multirow[c]{2}{*}{\bf Features}} &
    \multirow[c]{2}{*}{\bf Algorithm} &
    \multicolumn{2}{c|}{\bf Balanced accuracy}   \\
\cline{4-5}  
 & & & \bf \textit{Dataset-1} & \bf \textit{Dataset-2}\\ 
    \hline
\multirow[c]{28}{*}{CFG}           
            & Graph features \&          &  RandomForest & 0.65 & 0.57 \\  \cline{3-5} 
            & assembly (Dim: {\bf \#23}) &  GradientBoosting & 0.66 & 0.594 \\  \cline{2-5}
            & TF-IDF on assembly          & RandomForest & 0.697 & 0.593 \\  \cline{3-5} 
            & mnemonics (Dim: {\bf \#128})& GradientBoosting & 0.724 & 0.579 \\  \cline{2-5} 
            & \multirow[c]{5}{*}{Identity (Dim: {\bf \#1})}
            		&  GCN & 0.323  & 0.326 \\  \cline{3-5}
            &           &  Sage & 0.341 & 0.347   \\  \cline{3-5}
            &           &  GIN & 0.414 & 0.407 \\  \cline{3-5}
            &           &  GAT & 0.192 & 0.195      \\  \cline{3-5}
            &           &  UNet & 0.362 & 0.299      \\ \cline{2-5}
            & \multirow[b]{3}{*}{Counting mnemonic}
            		&  GCN & 0.431 & 0.462   \\  \cline{3-5}
            &           &  Sage & 0.498 & 0.499       \\  \cline{3-5}
            &           &  GIN & 0.488 & 0.474        \\  \cline{3-5}
            & classes (Dim: {\bf \#27}) &  GAT & 0.45 & 0.342      \\  \cline{3-5}
            &           &  UNet & 0.439 & 0.448   \\ \cline{2-5}
            &
                                  &  GCN & 0.721 & 0.675      \\  \cline{3-5}
            & Semantic \& counting&  Sage & 0.737 & 0.549    \\  \cline{3-5}
            & PCode mnemonics     &  GIN & 0.732 & 0.657        \\  \cline{3-5}
            & (Dim: {\bf \#78})   &  GAT & 0.729 & 0.637       \\  \cline{3-5}
            &                     &  UNet & 0.704 & 0.655        \\ \cline{2-5}
            &
                                  &  GCN & 0.723 & 0.633             \\  \cline{3-5}
            & Semantic \& counting&  Sage & 0.718 & 0.535            \\  \cline{3-5}
            & assembly mnemonics  &  GIN & 0.713 & 0.427             \\  \cline{3-5}
            & (Dim: {\bf \#1839}) &  GAT & 0.723 & 0.646          \\  \cline{3-5}
            &                     &  UNet  & 0.709 & 0.611        \\ \cline{1-5}
\end{tabular}
\label{multi_classif_O0}
\end{table}

\section{Multi-class classification} \label{sec:mul}
In multi-class classification, the goal is to determine the type of
obfuscation that has been applied to a function. Results for -O0 are available in Table
\ref{multi_classif_O0}. Many observations
from the binary case are confirmed in this experiment. Indeed,
elementary baselines perform remarkably well, given the fact there
exists 11 classes. 
%Combining GradientBoosting with TF-IDF features
%reaches 0.72 of balanced accuracy for the Dataset-1 and almost 0.60
%for the Dataset-2. 
As a comparison, previous works consider at most 4
classes \cite{Jiang2021GCNLSTM}. Similarly,
\glspl*{gnn} outperform baselines, especially on \textit{Dataset-2}, when the
features are enriched with semantic information originated from Pcode mnemonic. %\gls*{gnn} results with assembly mnemonic counts behave identically to the binary case. %In contrast, the assembly mnemonic counts feature shows slightly lower scores compared to the Pcode feature in the binary case on the \textit{Dataset-2}. 

\section{Real-world example: XTunnel} \label{sec:xtunnel}

%\subsection{XTunnel}
\begin{table}[t]
\caption{Obfuscation detection results on two XTunnel samples.}
\centering
\begin{tabular}{c|c|c}
	 & Binary balanced accuracy & Multi-class balanced accuracy \\
\hline
Sample C637E & 0.726 & 0.533 \\
Sample 99B45 & 0.711 & 0.55 \\
\end{tabular}
\label{xtunnel_exp}
\end{table}
XTunnel is a malware, developed by APT28 hacking group, that can relay traffic between a victim and a server used by cybercriminals to control compromised devices and exfiltrate data. Multiple variants have been found on governmental and institutional networks, for which some of them were obfuscated. This obfuscation has been used to evade security products. Malware deobfuscation helps to highlight and determine malicious functionalities hidden inside these obfuscated executables \cite{David2017XTunnel}. Then, locating and determining the obfuscation type is necessary before any deobfuscation attempt. These tasks are performed on two XTunnel obfuscated samples\footnote{Their corresponding hashes are C637E01F50F5FBD2160B191F6371C5DE2AC56DE4 and 99B454262DC26B081600E844371982A49D334E5E.}. Results are compared with a previously built ground truth, that was computed using a costly approach of symbolic execution \cite{David2017XTunnel}. Such ground truth asserts with a satisfactory confidence that these samples were heavily obfuscated using OpaquePredicates. The binary and multi-class classification are handled by the model that, on averaged, gives the best results, which is GIN with Pcode-based feature. The binary scores are computed over all the executable functions, whereas the multi-class scores are limited to obfuscated functions only. %as it is trained to characterize obfuscated functions. 
Results are available in Table \ref{xtunnel_exp} and show the validity of our approach. Interestingly, many functions are considered obfuscated with EncodeArithmetic instead of OpaquePredicates. Indeed, distinguishing these two obfuscations is tedious as an OpaquePredicate is simply an EncodeArithmetic followed by a branching instruction, with one fake branching. Consequently, our models can still detect a suspicious pattern related to OpaquePredicates and we consider these predictions as valid. 

%\subsection{Unc0ver}

%Unc0ver is an jailbreak tool dedicated to iOS, from iOS 11.0 to 14.8. It is used in order to obtain a complete access to a device and its system. Unc0ver is benign but performs offensive actions in order to jailbreak an iOS-based device. Such exploit is probably based on a 0-day vulnerability. In order to ensure this vulnerability cannot be used by malicious actors or patched by Apple, obfuscation is used to guarantee the Unc0ver sustainability. Similarly to the XTunnel example, we perform binary and multi-class classification on one Unc0ver sample, compiled for Arm64. This sample is almost 9 times bigger than XTunnel. We use the same GIN model than before with a Pcode-based initial feature. Results are compared with a manually extracted ground-truth, that highlights an extensive use of \gls*{cff}. Balanced accuracies for binary and multi-class classification are respectively \textcolor{red}{TODO}. Notice that this ground-truth may be prone to human bias and eventually contains mistakes as it is usually particularly challenging to ensure with complete certainty a function have been obfuscated. 

\section{Conclusion} \label{conclusion}

To conclude, this work provides a general study about obfuscation detection, at both binary and multi-class levels. It demonstrates the efficiency of standard baselines and most importantly the potential of \glspl*{gnn} that, combined with meaningful features conveying part of function semantics, achieve the best results. These results are confirmed with a real-world malware example. 

If this work seeks to be as complete as possible, it is subject to specific limitations. First, building a real-word obfuscated dataset implies a lot of implementation constraints. We try our best to represent the large variety of obfuscators and obfuscations, given accessible resources. Because obfuscation and optimizations are intertwined, it is difficult to ensure that the obfuscation was correctly applied and that the compiler optimizations do not remove or attenuate initial obfuscation, especially in -O2. As a result, our dataset may contain specific functions that differ from what they should be. Second, \gls*{gnn} hyperparameters were obtained with a budget constraint. As a consequence, specific \gls*{gnn}s may have been advantaged compared to others. As an example, GAT takes a long time to train compared to simpler models such as GCN. 

Finally, this works constitutes only a first step of a more general study on obfuscation detection. More attention should be dedicated to innovating graph types and features that should capture as much as possible the function semantics. The binary similarity problem \cite{Marcelli2022Usenix} faces the same challenge, leading to the development of new graphs, such as \glsfmtfull{sog} \cite{He2024SOG} that is, to the best of our knowledge, the first attempt that tries to represent binary code by combining multiple edge types (data, control-flow, effects) inside a graph using solely disassembly. This representation seems promising as it brings together all the key aspects of a function, in particular part of its semantics.

\section*{Acknowledgments}
The authors thank the Agence Innovation Defense (AID) for its financial support.

\bibliographystyle{spmpsci} % We choose the "plain" reference style
\bibliography{refs} % Entries are in the refs.bib file

\begin{thebibliography}{10}
\providecommand{\url}[1]{{#1}}
\providecommand{\urlprefix}{URL }
\expandafter\ifx\csname urlstyle\endcsname\relax
  \providecommand{\doi}[1]{DOI~\discretionary{}{}{}#1}\else
  \providecommand{\doi}{DOI~\discretionary{}{}{}\begingroup
  \urlstyle{rm}\Url}\fi

\bibitem{optuna}
Akiba, T., Sano, S., Yanase, T., Ohta, T., Koyama, M.: Optuna: A
  next-generation hyperparameter optimization framework.
\newblock In: Proceedings of the 25th ACM SIGKDD international conference on
  knowledge discovery \& data mining, pp. 2623--2631 (2019)

\bibitem{David2017XTunnel}
Bardin, S., David, R., Marion, J.Y.: Backward-bounded dse: Targeting
  infeasibility questions on obfuscated codes.
\newblock In: 2017 IEEE Symposium on Security and Privacy (SP), pp. 633--651
  (2017).
\newblock \doi{10.1109/SP.2017.36}

\bibitem{tigress}
Collberg, C.: The tigress c obfuscator.
\newblock \url{https://tigress.wtf/index.html}.
\newblock Accessed: 2023-08-17

\bibitem{David2020DeobfuscationMBA}
David, R., Coniglio, L., Ceccato, M.: Qsynth-a program synthesis based approach
  for binary code deobfuscation.
\newblock In: BAR 2020 Workshop (2020)

\bibitem{Errica2020A}
Errica, F., Podda, M., Bacciu, D., Micheli, A.: A fair comparison of graph
  neural networks for graph classification.
\newblock In: International Conference on Learning Representations (2020).
\newblock \urlprefix\url{https://openreview.net/forum?id=HygDF6NFPB}

\bibitem{Ferrari2019Progress}
Ferrari~Dacrema, M., Cremonesi, P., Jannach, D.: Are we really making much
  progress? a worrying analysis of recent neural recommendation approaches.
\newblock In: Proceedings of the 13th ACM Conference on Recommender Systems,
  RecSys ’19. ACM (2019).
\newblock \doi{10.1145/3298689.3347058}.
\newblock \urlprefix\url{http://dx.doi.org/10.1145/3298689.3347058}

\bibitem{torchgeometric}
Fey, M., Lenssen, J.E.: Fast graph representation learning with {PyTorch
  Geometric}.
\newblock In: ICLR Workshop on Representation Learning on Graphs and Manifolds
  (2019)

\bibitem{Gao2019Unet}
Gao, H., Ji, S.: Graph u-nets.
\newblock In: K.~Chaudhuri, R.~Salakhutdinov (eds.) Proceedings of the 36th
  International Conference on Machine Learning, \emph{Proceedings of Machine
  Learning Research}, vol.~97, pp. 2083--2092. PMLR (2019).
\newblock \urlprefix\url{https://proceedings.mlr.press/v97/gao19a.html}

\bibitem{Gori2005GNN}
Gori, M., Monfardini, G., Scarselli, F.: A new model for learning in graph
  domains.
\newblock In: Proceedings. 2005 IEEE International Joint Conference on Neural
  Networks, 2005., vol.~2, pp. 729--734 vol. 2 (2005).
\newblock \doi{10.1109/IJCNN.2005.1555942}

\bibitem{Greco2023Explaining}
Greco, C., Ianni, M., Guzzo, A., Fortino, G.: Explaining binary obfuscation.
\newblock pp. 22--27 (2023).
\newblock \doi{10.1109/CSR57506.2023.10224825}

\bibitem{Hamilton2018SAGE}
Hamilton, W.L., Ying, R., Leskovec, J.: Inductive representation learning on
  large graphs.
\newblock In: Proceedings of the 31st International Conference on Neural
  Information Processing Systems, NIPS'17, p. 1025–1035. Curran Associates
  Inc., Red Hook, NY, USA (2017)

\bibitem{He2024SOG}
He, H., Lin, X., Weng, Z., Zhao, R., Gan, S., Chen, L., Ji, Y., Wang, J., Xue,
  Z.: Code is not natural language: Unlock the power of semantics-oriented
  graph representation for binary code similarity detection.
\newblock In: 33rd USENIX Security Symposium (USENIX Security 24),
  PHILADELPHIA, PA (2024)

\bibitem{Jiang2021GCNLSTM}
Jiang, S., Hong, Y., Fu, C., Qian, Y., Han, L.: Function-level obfuscation
  detection method based on graph convolutional networks.
\newblock Journal of Information Security and Applications \textbf{61}, 102,953
  (2021).
\newblock \doi{https://doi.org/10.1016/j.jisa.2021.102953}.
\newblock
  \urlprefix\url{https://www.sciencedirect.com/science/article/pii/S2214212621001654}

\bibitem{ollvm}
Junod, P., Rinaldini, J., Wehrli, J., Michielin, J.: Obfuscator-llvm--software
  protection for the masses.
\newblock In: B.~Wyseur (ed.) Proceedings of the IEEE/ACM 1st International
  Workshop on Software Protection, SPRO'15, Firenze, Italy, May 19th, 2015, pp.
  3--9. IEEE (2015).
\newblock \doi{10.1109/SPRO.2015.10}

\bibitem{Kanzaki2015Artificiality}
Kanzaki, Y., Monden, A., Collberg, C.: Code artificiality: A metric for the
  code stealth based on an n-gram model.
\newblock In: 2015 IEEE/ACM 1st International Workshop on Software Protection,
  pp. 31--37. IEEE (2015)

\bibitem{Welling2017GCN}
Kipf, T.N., Welling, M.: Semi-supervised classification with graph
  convolutional networks.
\newblock In: International Conference on Learning Representations (2017).
\newblock \urlprefix\url{https://openreview.net/forum?id=SJU4ayYgl}

\bibitem{Kriege2020}
Kriege, N.M., Johansson, F.D., Morris, C.: A survey on graph kernels.
\newblock Applied Network Science \textbf{5}(1), 6 (2020).
\newblock \doi{10.1007/s41109-019-0195-3}.
\newblock \urlprefix\url{https://doi.org/10.1007/s41109-019-0195-3}

\bibitem{Marcelli2022Usenix}
Marcelli, A., Graziano, M., Ugarte-Pedrero, X., Fratantonio, Y., Mansouri, M.,
  Balzarotti, D.: How machine learning is solving the binary function
  similarity problem.
\newblock In: 31st USENIX Security Symposium (USENIX Security 22), pp.
  2099--2116 (2022)

\bibitem{Collberg2009Book}
Nagra, J., Collberg, C.: Surreptitious Software: Obfuscation, Watermarking, and
  Tamperproofing for Software Protection: Obfuscation, Watermarking, and
  Tamperproofing for Software Protection.
\newblock Pearson Education (2009)

\bibitem{scikit_learn}
Pedregosa, F., Varoquaux, G., Gramfort, A., Michel, V., Thirion, B., Grisel,
  O., Blondel, M., Prettenhofer, P., Weiss, R., Dubourg, V., et~al.:
  Scikit-learn: Machine learning in python.
\newblock the Journal of machine Learning research \textbf{12}, 2825--2830
  (2011)

\bibitem{obfdataset}
Quarkslab: Obfuscation dataset.
\newblock \url{https://github.com/quarkslab/diffing_obfuscation_dataset}.
\newblock Accessed: 2024-09-01

\bibitem{Raja2007RE}
Raja, V., Fernandes, K.J.: Reverse engineering: an industrial perspective.
\newblock Springer Science \& Business Media (2007)

\bibitem{Salem2016TFIDF}
Salem, A., Banescu, S.: Metadata recovery from obfuscated programs using
  machine learning.
\newblock In: Proceedings of the 6th Workshop on Software Security, Protection,
  and Reverse Engineering, pp. 1--11 (2016)

\bibitem{Salwan2018DeobfuVirtualize}
Salwan, J., Bardin, S., Potet, M.L.: Symbolic deobfuscation: From virtualized
  code back to the original.
\newblock In: International Conference on Detection of Intrusions and Malware,
  and Vulnerability Assessment, pp. 372--392. Springer (2018)

\bibitem{Schrittwieser2023Stealth}
Schrittwieser, S., Wimmer, E., Mallinger, K., Kochberger, P., Lawitschka, C.,
  Raubitzek, S., Weippl, E.R.: Modeling obfuscation stealth through code
  complexity.
\newblock In: European Symposium on Research in Computer Security, pp.
  392--408. Springer (2023)

\bibitem{Sharif2008Malware}
Sharif, M.I., Lanzi, A., Giffin, J.T., Lee, W.: Impeding malware analysis using
  conditional code obfuscation.
\newblock In: NDSS (2008)

\bibitem{Ramtine2019Opaque}
Tofighi-Shirazi, R., Asavoae, I.M., Elbaz-Vincent, P., Le, T.H.: Defeating
  opaque predicates statically through machine learning and binary analysis.
\newblock In: Proceedings of the 3rd ACM Workshop on Software Protection, pp.
  3--14 (2019)

\bibitem{Ramtine2019Stealth}
Tofighi-Shirazi, R., As\u{a}voae, I.M., Elbaz-Vincent, P.: Fine-grained static
  detection of obfuscation transforms using ensemble-learning and semantic
  reasoning.
\newblock In: Proceedings of the 9th Workshop on Software Security, Protection,
  and Reverse Engineering, SSPREW9 '19. Association for Computing Machinery,
  New York, NY, USA (2019).
\newblock \doi{10.1145/3371307.3371313}.
\newblock \urlprefix\url{https://doi.org/10.1145/3371307.3371313}

\bibitem{Velickovic2018GAT}
Veličković, P., Cucurull, G., Casanova, A., Romero, A., Liò, P., Bengio, Y.:
  Graph attention networks.
\newblock In: International Conference on Learning Representations (2018).
\newblock \urlprefix\url{https://openreview.net/forum?id=rJXMpikCZ}

\bibitem{Wang2001CFFDef}
Wang, C.: A security architecture for survivability mechanisms.
\newblock University of Virginia (2001)

\bibitem{Xu2019Howpowerful}
Xu, K., Hu, W., Leskovec, J., Jegelka, S.: How powerful are graph neural
  networks?
\newblock In: International Conference on Learning Representations (2019).
\newblock \urlprefix\url{https://openreview.net/forum?id=ryGs6iA5Km}

\bibitem{Yadegari2015DeobfuscationGeneral}
Yadegari, B., Johannesmeyer, B., Whitely, B., Debray, S.: A generic approach to
  automatic deobfuscation of executable code.
\newblock In: 2015 IEEE Symposium on Security and Privacy, pp. 674--691 (2015).
\newblock \doi{10.1109/SP.2015.47}

\bibitem{Zhou2007MBADef}
Zhou, Y., Main, A., Gu, Y.X., Johnson, H.: Information hiding in software with
  mixed boolean-arithmetic transforms.
\newblock In: International Workshop on Information Security Applications, pp.
  61--75. Springer (2007)

\end{thebibliography}
\end{document}